# HIGH TEMPERATURE NEUTRON DIFFRACTION STUDY OF THE $La_{1.4}Sr_{1.6}Mn_2O_7$ BILAYERED MANGANITE

LORENZO MALAVASI[a,*], GAETANO CHIODELLI[a], HÅKAN RÜNDLOF[b], CRISTINA TEALDI[a], AND GIORGIO FLOR[a]

[a] *Dipartimento di Chimica Fisica "M. Rolla" of University of Pavia, INSTM, IENI/CNR Department of Pavia, V.le Taramelli 16, I-27100, Pavia, Italy.*

**E-mail: lorenzo.malavasi@unipv.it*

[b] *Studsvik Neutron Research Lab., Studsvik, S-611 82, Nykoping, Sweden*

## Abstract

In this paper we present the results of a high temperature (300 K≤$T$≤800 K) neutron powder diffraction study of the $La_{1.4}Sr_{1.6}Mn_2O_7$ bilayered manganite. An unusual trend of the <Mn-$O_{apical}$>/Mn-$O_{equatorial}$ parameter was found: it first decreases up to 500 K and then increases up to the highest $T$ measured. At the same time, a significant shortening of the apical Mn-O(2) bond is observed in the range where the J-T distortion is reduced. The overall data gained by this study may suggest a shift of electronic density from axial to planar $e_g$ orbitals with $T$. This trend is explained considering of the presence of short range magnetic interaction well above $T_C$.

There is an increasing interest in materials that show magnetoresistance, because of their use in magnetic information storage or as magnetic field sensors [1]. Most of the available theoretical and experimental work has, until now, been focused on the 3D perovskite structures, that is the $n = \infty$ end-member of the $A_{n+1}B_nO_{3n+1}$ Ruddlesden-Popper family, in which $n$ 2D layers of $BO_6$ corner-sharing octahedra are joined along the stacking direction and separated by rock-salt AO layers.

The optimally-doped $n = 2$ members of this family ($La_{2-2x}B_{1+2x}Mn_2O_7$ where B = Ca or Sr) behave analogously to the $n = \infty$ manganites in the sense that they undergo an insulating- to metallic-like state (I-M) transition coupled to a ferromagnetic transition at temperatures around 120-140 K; besides, they present a large magnetoresistance in this temperature range [2]. However, the change in dimensionality and resulting pronounced cation dependence of the electronic properties can produce physical properties which contrast strongly with the perovskite systems [3-6].

For example, the temperature-dependent structural properties of these anisotropic systems present remarkable differences with respect to the 3D ones. Mitchell and co-workers [7] showed, by means of neutron diffraction up to 500 K, that, unlike the related manganite perovskites, whose Jahn-Teller distorted octahedra become more regular for temperatures below the I-M transition, the $MnO_6$ octahedra of an optimally doped layered manganite ($La_{1.2}Sr_{1.8}Mn_2O_7$) are more severely distorted when the charge is itinerant than when it is localized. However, this study discussed mainly the low-$T$ range also because of the relatively small high temperature interval spanned.

To the best of our knowledge, the neutron diffraction study reported in [7] spanned the wider $T$-range on a bilayered manganite. However, we think that a neutron diffraction study with the aim to extend this temperature range towards the high-$T$ region would be of great importance. The main reason of such a study is to follow the J-T distortion well above the Curie temperature (which is around 140 K) in order to search for an analogous of the $T_{J\text{-}T}$ found for the 3D manganite and which corresponds to the total removal of the J-T distortion as a consequence, also, of the charge disproportionation of Mn(III) into Mn(II) and Mn(IV) non J-T ions [3]. In the stoichiometric $LaMnO_3$, by increasing the temperature, a rich structural evolution was found. The sample changes from a *orbitally* ordered orthorhombic structure (with cooperative and static J-T distortion) towards a less distorted orthorhombic structure characterized by a dynamic J-T distortion. Conclusions were derived, in that work, from both the evolution of the Mn-O bond lengths and from the oxygen thermal factors values as a function of *T*. For the $La_{1.2}Sr_{1.8}Mn_2O_7$ manganite, up to 500 K, the presence of a "*one-long, five short*" $MnO_6$ octahedron has been found. This was attributed not only to electronic but also to steric effects [7].

Our aim is then to follow the structural evolution of the layered manganite $La_{1.4}Sr_{1.6}Mn_2O_7$ in the 300 K$\leq T \leq$800 K range by means of neutron powder diffraction in order to be able to derive reliable atomic position parameters and thermal factors.

$La_{1.4}Sr_{1.6}Mn_2O_7$ was synthesized by solid state reaction starting from proper amounts of $La_2O_3$, $Mn_2O_3$, and $SrCO_3$ (Aldrich >99.99%). Pellets were prepared from the thoroughly mixed powders and allowed to react first at 1273 K for 72 hours and afterwards at 1573 K for 72 hours. During the thermal treatments pellets were re-ground and re-pelletized at least three times.

X-ray powder diffraction patterns (XRPD) were acquired on a Bruker "D8 Advance" instrument. X-ray diffraction indicated the single phase nature of the prepared material which can be perfectly indexed with a tetragonal unit cell within the *I*4/*mmm* space group (no. 139). High temperature neutron diffraction (ND) patterns were collected every 100 K from 300 K to 800 K on the NPD instrument at the Studsvik Nuclear Research Center (SE) by employing a wavelength of 1.47 Å. Measurements were collected in the 2θ-range 4-140° with 0.08° step size. Counting time was 12 min per step. Diffraction patterns were refined by means of Rietveld method with the FULLPROF software [8]. Sample chemical composition was checked by means of Electron Microprobe Analysis (EMPA) and it was found in agreement with the expected nominal one.

Magnetisation and electrical transport measurements confirmed the presence of magnetic and I-M transitions around 130 K. High-$T$ conductivity measurements were carried out with the DC-four probes method on a sample pellet.

Figure 1 shows the refined neutron pattern collected at 300 K for the $La_{1.4}Sr_{1.6}Mn_2O_7$ manganite (hereafter indicated as LSMO). As can be seen, the structural model fits very well with the experimental pattern. However, best fit was obtained by considering a two phase model where, beside the double-layered LSMO, also a perovskite manganite is taken into account (strongest peak of this phase is highlighted by an asterisks in Figure 1). Quantitative phase analysis through the refinement estimated the amount of the impurity phase to be ~4%.

Table 1 reports the structural parameters determined from the Rietveld analysis of neutron powder diffraction data collected at the six temperatures investigated. In all the refinements the cations occupancies were kept fixed to the nominal values, also according to the chemical analysis of the samples, while those of oxygen atoms where refined without any constrain. The line shape was modelled with a pseudo-Voigt profile and an isotropic thermal factor was refined for each atom.

Figure 2 shows the temperature evolution of the unit cell parameters $a$($b$) and $c$ with temperature. Both the lattice constants increase as the temperature is raised. However, the relative variation from 300 K to 800 K is ~0.5% for the $a$ lattice constant and ~0.9% for the $c$ constant. In the Figure inset we reported the variation rate of the lattice parameter with temperature. As can be appreciated, a strong anisotropy is present in the rate with which the constants increase with $T$, revealing an interesting trend: for $T \leq 500°C$ the short axis ($a$) increases faster then the long one ($c$) as $T$ is increased, while for $T > 500°C$ this behaviour is reversed.

Figure 3 shows the cell volume and $c/a$ variation with $T$. We stress that the $c/a$ parameter can be considered as a marker of the tetragonal distortion of the cell. It can be seen that is does not follow a linear behaviour with $T$ as, in contrast, is found for the cell volume. It is clear that this last parameters (V) reflects the overall $T$-dependence of the sample but, as clearly suggested by the other structural parameters, significant anisotropies are present among the two lattice constants. The origin of these must be seek in the trend of Mn-O bond lengths which, in turn, are influenced by strong effects of the electronic structure (*i.e.*: orbital occupancies of $e_g$ electrons and J-T distortions).

Figure 4 gives a more detailed description of the Mn-O bond lengths fluctuations as a function of temperature. The most noticeable behaviour is the contraction of the Mn-O(2) apical bond length as $T$ is increases up to 500 K. The other two bonds expand roughly linearly between RT and 500 K. Above this threshold (marked with a vertical dashed line in Figure 4) a different behaviour occurs for the bond lengths. Now, the apical Mn-O(2) bond starts expanding significantly with $T$ while the other apical bond, that is the bond to the oxygen lying between the $MnO_2$ bilayers, shows a sort of plateau ended with a significant rise going to 800 K.

Temperature dependence of the three symmetry independent Mn-O bonds can be directly correlated to the distortion of the $MnO_6$ octahedra and consequently to the Jahn-Teller distortion in the $La_{1.4}Sr_{1.6}Mn_2O_7$ layered manganite. For this purpose, the J-T distortion parameter $D$ (defined as $D = \langle Mn\text{-}O_{apical} \rangle / Mn\text{-}O_{equatorial}$) is shown in Figure 5 as a function of temperature.

Noticeably, the $D$ parameter decreases with decreasing temperature from 300 to 500 K and then start increasing up to the highest temperature measured (800 K). This result correlates directly to the bond lengths trend observed in Figure 4. From 300 K to 500 K the contraction of the Mn-O(2) apical bond together with the expansion of the apical Mn-O(1) and equatorial Mn-O(3) bonds induces a reduction of the J-T distortion which then progressively increases by raising the temperature. This is opposite to what observed in the $LaMnO_3$ perovskite where a progressive reduction of the J-T distortion by increasing the temperature was found [9].

Finally, Figure 6 shows the electrical conductivity measured for the $La_{1.4}Sr_{1.6}Mn_2O_7$ manganite in the same $T$-range of the neutron diffraction measurements. We successfully fit the curve with the classical semiconducting law $\rho = \rho_0 \exp(-E_a / k_B T)$. Careful inspection of the curve revealed a small slope change around 520-530 K. Activation energies ($E_a$) from RT to about 520 K is around 43 meV while from about 520 K to 800 K it is slightly reduced to 39.5 meV. We remark that a reasonable fit of the experimental data might be also obtained by a single exponential curve but due to the observed anomalies in the diffraction data around 500 K we tried to look if any marker of that behaviour was present on the electrical transport data.

In order to try to understand the unusual behaviour found at high temperature for the $La_{1.4}Sr_{1.6}Mn_2O_7$ manganite we have to recall the basic magnetic and transport properties of this compound. Tokura and co-workers have shown [2-3] that the phase diagram of Sr-doped bilayered manganites is strongly doping-dependent. At $x$ = 0.3, the magnetic moments of each $MnO_2$ layer couple ferromagnetically within a bilayer and antiferromagnetically, along the $c$-axis, between successive bilayers. Moreover, it was observed that a two-dimensional ferromagnetic short-range interaction exists in a wide temperature region above $T_C$ which may be related to the anisotropic exchange energy ($|J_{ab}| > |J_c| >> |J'|$, $J_{ab}$, $J_c$, and $J'$ stand for the in-plane, inter-single-layer, and inter-bilayer exchange interaction, respectively) [2-3]. The strong magnetic anisotropy that characterizes these layered systems also affects the transport properties due to the close coupling between magnetism and electronic structure. Single-crystals studies have shown that, in the $0.3 \leq x \leq 0.4$ range, for $La_{2-2x}Sr_{1+2x}Mn_2O_7$, the $\rho_c/\rho_{ab}$ (where $\rho_c$ and $\rho_{ab}$ are the resistivity along the $c$-axis and in the $a$-$b$ plane, respectively) is as large as $10^2$ at room temperature which suggests a confinement of the carrier motion within the $MnO_2$ bilayer. Usually, the on-set of the long-range FM order is accompanied by a resistivity drop (the insulator-to-metal transition) followed by a metallic-like transport. In the $T$-range above $T_C$ both $\rho_c$ and $\rho_{ab}$ show an activated-like transport with hopping energies around 30-40 meV [10].

Conductivity anisotropies between the in-plane and out of plane transport are usually observed up to temperatures very far from the 3D magnetic order. Velázquez and co-workers [11] showed that, for the $La_{1.2}Sr_{1.8}Mn_2O_7$ manganite ($x$ = 0.4), the $\rho_c/\rho_{ab}$ ratio becomes constant at temperatures higher than ~425 K. A constant value of the $\rho_c/\rho_{ab}$ ratio (even though different from 1) indicates the absence of any significant spin correlations along particular directions (*a-b* plane or *c*-axis) and it is in agreement with the observed mean-field magnetic behaviour in all directions [12].

The progressive reduction of the *D* parameter from 300 K to 500 K and the progressive shortening of the apical Mn-O(2) bond as temperature is raised can not be accounted for by only considering thermal structural effects. These results points out that also an electronic effect should play a role which is also indicative of some spin correlation still present at this temperature.

The $x = 0.3$ sample is characterized by a preferential $3z^2$-$r^2$ $e_g$ orbital occupancy which is reflected in a severe J-T distorted structure with significant long Mn-O(2) bond [3]. It seems that by increasing the temperature significant electronic fluctuations take place. In particular a progressive shift of electronic density from the apical orbitals to the planar $3x^2$-$y^2$ states should occur. This will lead to a overall reduction of the J-T distortion as witnessed by the *D* parameter reduction. Also the faster variation of the in-plane lattice parameter (see Figure 2) moving towards 500 K may be explained by considering a significant increase of electronic density within the planes.

In some way this behaviour resembles the one found for the $La_{2-2x}Sr_{1+2x}Mn_2O_7$ solid solution as a function of doping but, obviously, in the present case no hole doping variation occurs. We also note that in order to allow this variation in the electronic distribution some changes would also occur in the magnetic interaction. We may consider that, according also to the high-temperature resistivity [11] and our EPR data [13], spin correlations are significant also in a *T*-range above RT. Increasing *T* increases the spin fluctuation but the energy gained may also trigger a change of the spin direction thus leading to a preferential *in-plane* short-range correlation opposite to what observed at low-*T*, where *c*-axis orientation is found, as a consequence of the enhancement of the $J_c$.

Above 500 K no more correlation are present and the system behaves as a paramagnetic semiconductor showing an overall expansion of all the three Mn-O bonds. The progressive increase of the *D* parameter indicates that the J-T distortion increases with the charge localization. However, a smaller activation energy was found for $T > 500$ K. This correlates to the difference in the band dispersion as a function of the $e_g$ electrons orbital character. As the population of the $3z^2$-$r^2$ $e_g$ orbital is increased a wider band is observed with respect to the situation for $T < 500$ K where some degree of preferential occupancy of the planar $3x^2$-$y^2$ states would induce a slight bandwidth reduction [2].

We finally remark that our results can not be directly compared to the available neutron powder diffraction data mainly, because of the reduced high-$T$ range studied by other Authors (never higher than 500 K). In addition, all the other works dealt with the $La_{1.2}Sr_{1.8}Mn_2O_7$ compound, where a different spin arrangement is found relative to the $x = 0.3$ composition. However, the fact that no anomalies in the structural parameter have been encountered on the $La_{1.2}Sr_{1.8}Mn_2O_7$ manganite (from 10 to 500 K) may be a further evidence of our findings since this composition already posses an in-plane spin arrangement with electron orbital occupancy mainly in the planar $3x^2$-$y^2$ states. Absence of any anomaly may also suggest, indirectly, that preferred spin arrangement lies, for bilayered manganites, within the planes instead of along the $c$-direction.

This work has reported the first neutron diffraction structural study on an optimally doped bilayered manganite, $La_{1.4}Sr_{1.6}Mn_2O_7$, for temperatures up to 800 K. An unusual trend has been found for the J-T distortion parameter ($D$) which first decreases from 300 to 500 K and then increases up to the highest $T$ measured. A the same time, a strong reduction of the apical Mn-O(2) bond is found in the range where J-T is reduced. The overall data gained by this study may suggest a shift of electronic density from axial to planar $e_g$ orbitals with $T$. We argue that short-range magnetic correlation are still present for temperatures greater than RT and play a role in the observed behaviour.

## Acknowledgements

Financial support from the Italian Ministry of University (MIUR) through the PRIN 2004 projects is gratefully acknowledged. NFL neutron facility and European Community financial support is acknowledged.

## Table Caption

**Table 1** – Structural parameters and agreement indices of the Rietveld refinements of the neutron powder diffraction patterns for $La_{1.2}Sr_{1.8}Mn_2O_7$ in the 300 K≤$T$≤800 K range.

## Figures Caption

**Figure 1** – Rietveld refined pattern at 300 K for $La_{1.2}Sr_{1.8}Mn_2O_7$. Red circles: experimental points; black line: calculated pattern; blue line: residues; vertical green lines: Bragg peaks of the I/4*mmm* tetragonal structure.

**Figure 2** – *a* (full squares) and *c* (empty squares) lattice parameters variation as a function of temperature. Inset: derivative of *a* (full squares) and *c* (empty squares) lattice parameters with respect to *T*. Lines are guides for the eyes only.

**Figure 3** – Cell volume (full squares) and *c*/*a* (empty squares) variation as a function of temperature. Lines are guides for the eyes only.

**Figure 4** – Mn-O bond lengths behaviour as a function of temperature. See legend in the Figure for details. Lines are guides for the eyes only.

**Figure 5** – $\langle Mn\text{-}O_{apical}\rangle / Mn\text{-}O_{equatorial}$ parameter variation as a function of temperature.

**Figure 6** – Resistivity as a function of temperature. Inset: Logarithm of resistivity *vs*. 1/*T*.

# Table 1

| | | 295 K | 400 K | 500 K | 600 K | 700 K | 800 K |
|---|---|---|---|---|---|---|---|
| | *a* (Å) | 3.8719(1) | 3.8756(1) | 3.8799(1) | 3.8838(1) | 3.8878(2) | 3.8915(2) |
| | *c* (Å) | 20.258(2) | 20.292(1) | 20.322(1) | 20.355(1) | 20.394(1) | 20.441(1) |
| | V ($Å^3$) | 303.71(2) | 304.81(2) | 305.92(2) | 307.05(2) | 308.26(2) | 309.56(2) |
| **La/Sr (1)** | *B* | 0.32(7) | 0.44(9) | 0.54(9) | 0.8(1) | 0.9(1) | 1.2(1) |
| **La/Sr (2)** | *z* | 0.3176(2) | 0.3175(2) | 0.3174(2) | 0.3174(2) | 0.3176(2) | 0.3175(2) |
| | *B* | 0.37(5) | 0.63(6) | 0.85(6) | 1.03(6) | 1.22(7) | 1.27(7) |
| **Mn** | *z* | 0.0964(3) | 0.0965(3) | 0.0966(3) | 0.0964(3) | 0.0962(3) | 0.0962(3) |
| | *B* | 0.48(7) | 0.49(7) | 0.62(8) | 0.61(8) | 0.71(8) | 0.76(9) |
| **O(1)** | *B* | 0.6(1) | 1.2(1) | 1.4(1) | 1.5(2) | 1.7(2) | 2.0(2) |
| **O(2)** | *z* | 0.1970(2) | 0.1967(2) | 0.1964(2) | 0.1969(2) | 0.1968(2) | 0.1971(2) |
| | *B* | 1.0(1) | 1.1(1) | 1.3(1) | 1.5(1) | 1.5(1) | 2.0(2) |
| **O(3)** | *z* | 0.0955(1) | 0.0954(1) | 0.0952(1) | 0.0953(2) | 0.0955(2) | 0.0955(2) |
| | *B* | 0.91(6) | 1.16(7) | 1.34(7) | 1.46(7) | 1.79(8) | 1.85(8) |
| $R_B$ | | 2.87 | 2.61 | 2.91 | 3.30 | 3.53 | 3.22 |
| $R_{wp}$ | | 5.18 | 5.19 | 4.96 | 5.12 | 5.23 | 5.03 |

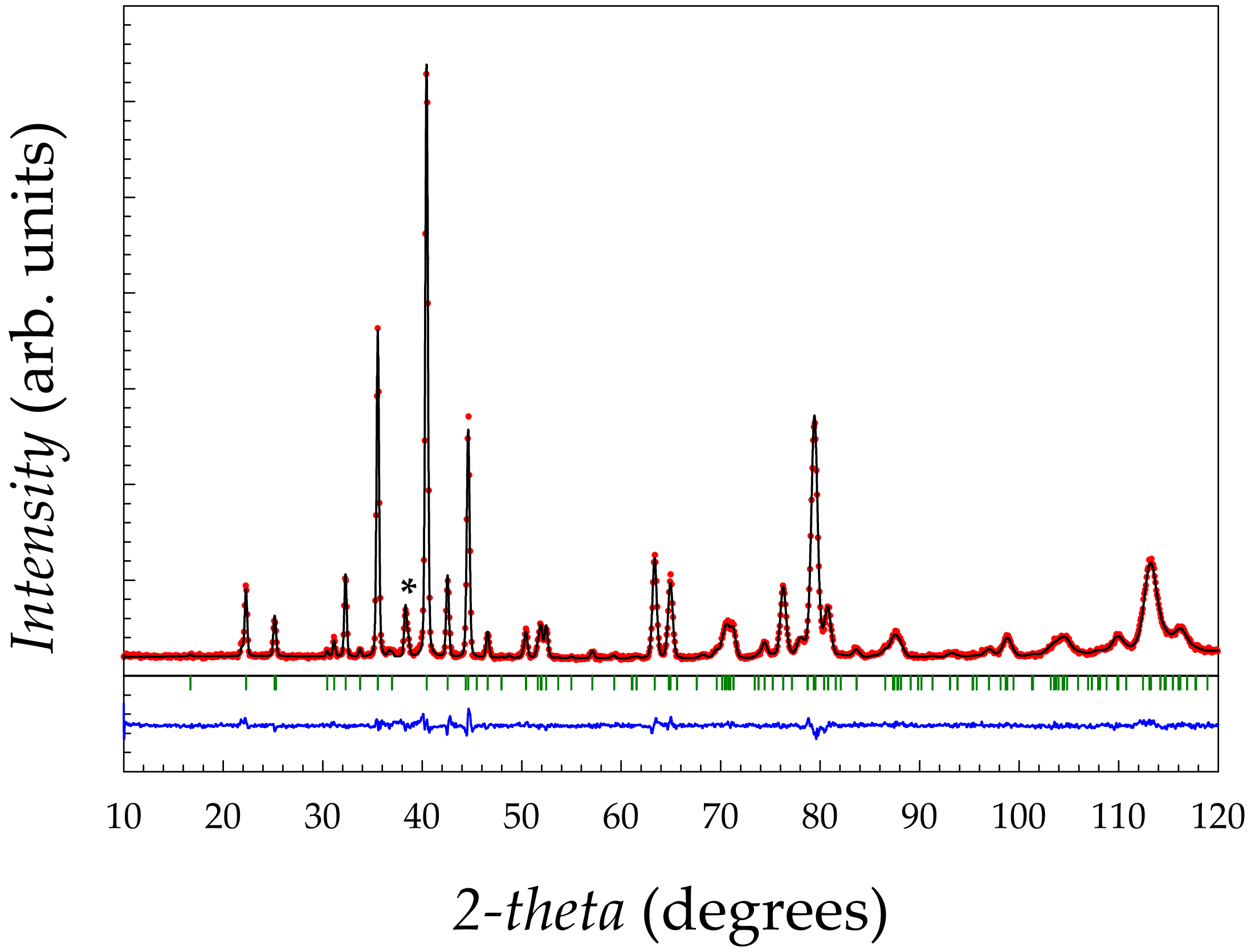

**Figure 1**

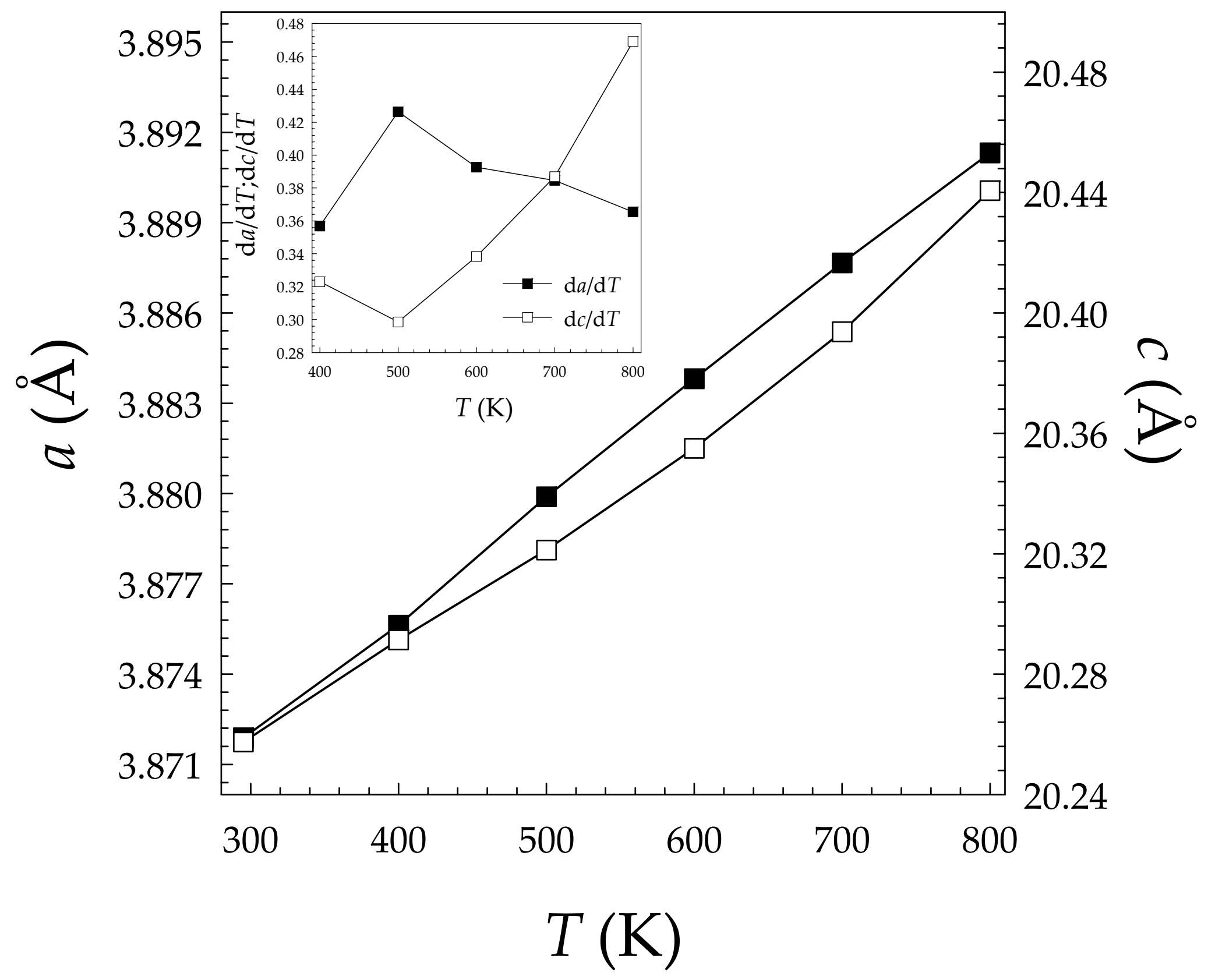

**Figure 2**

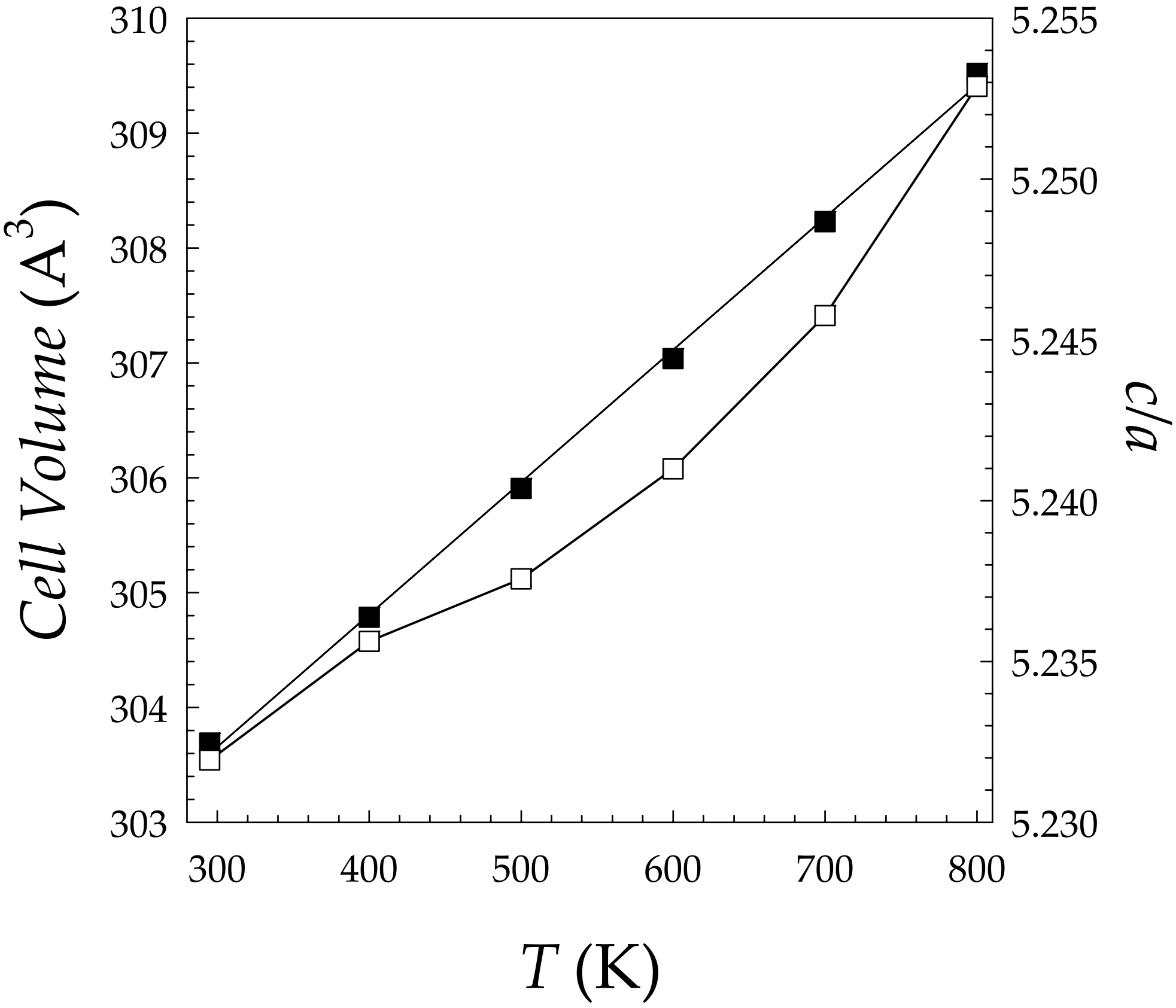

**Figure 3**

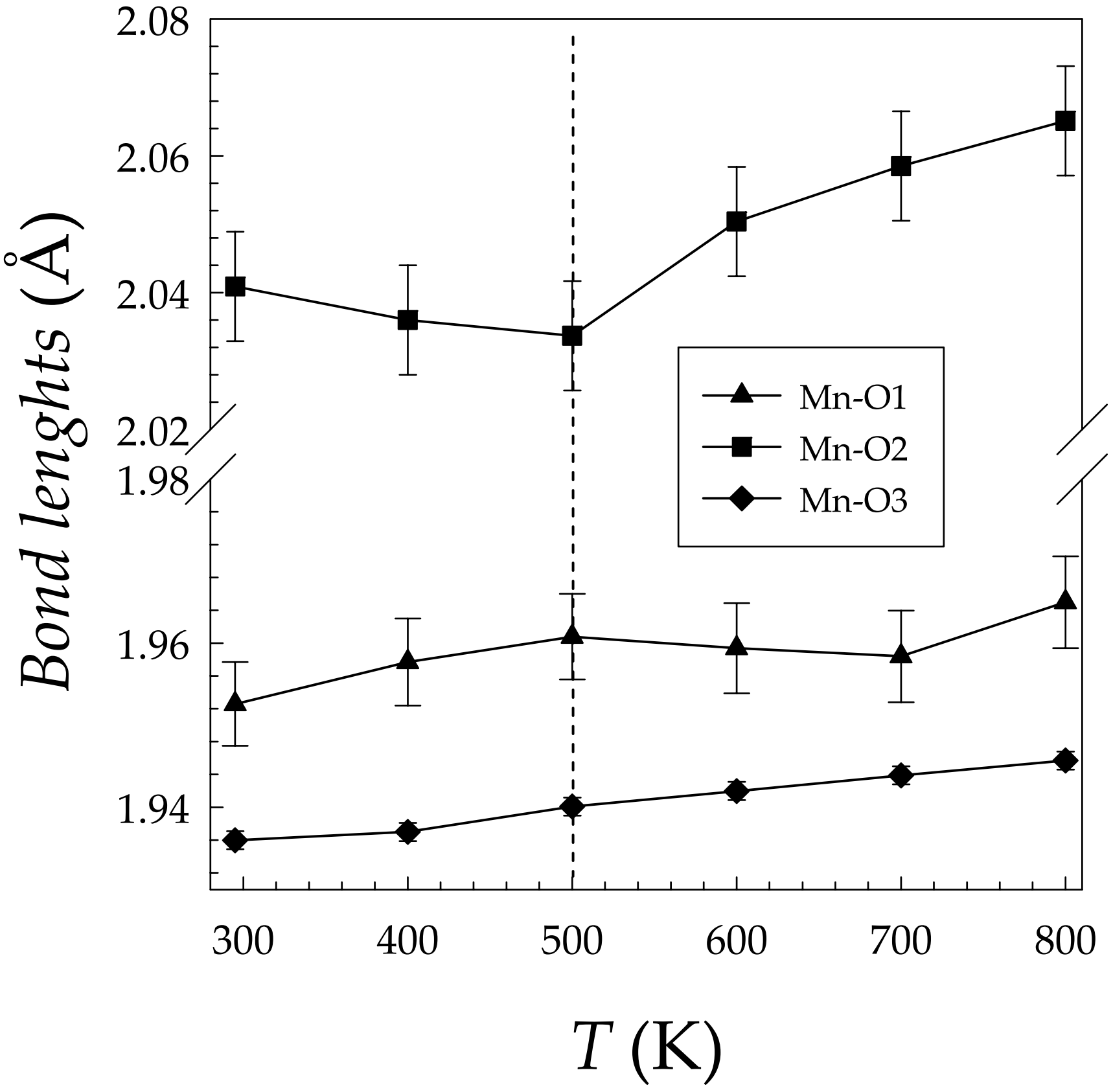

**Figure 4**

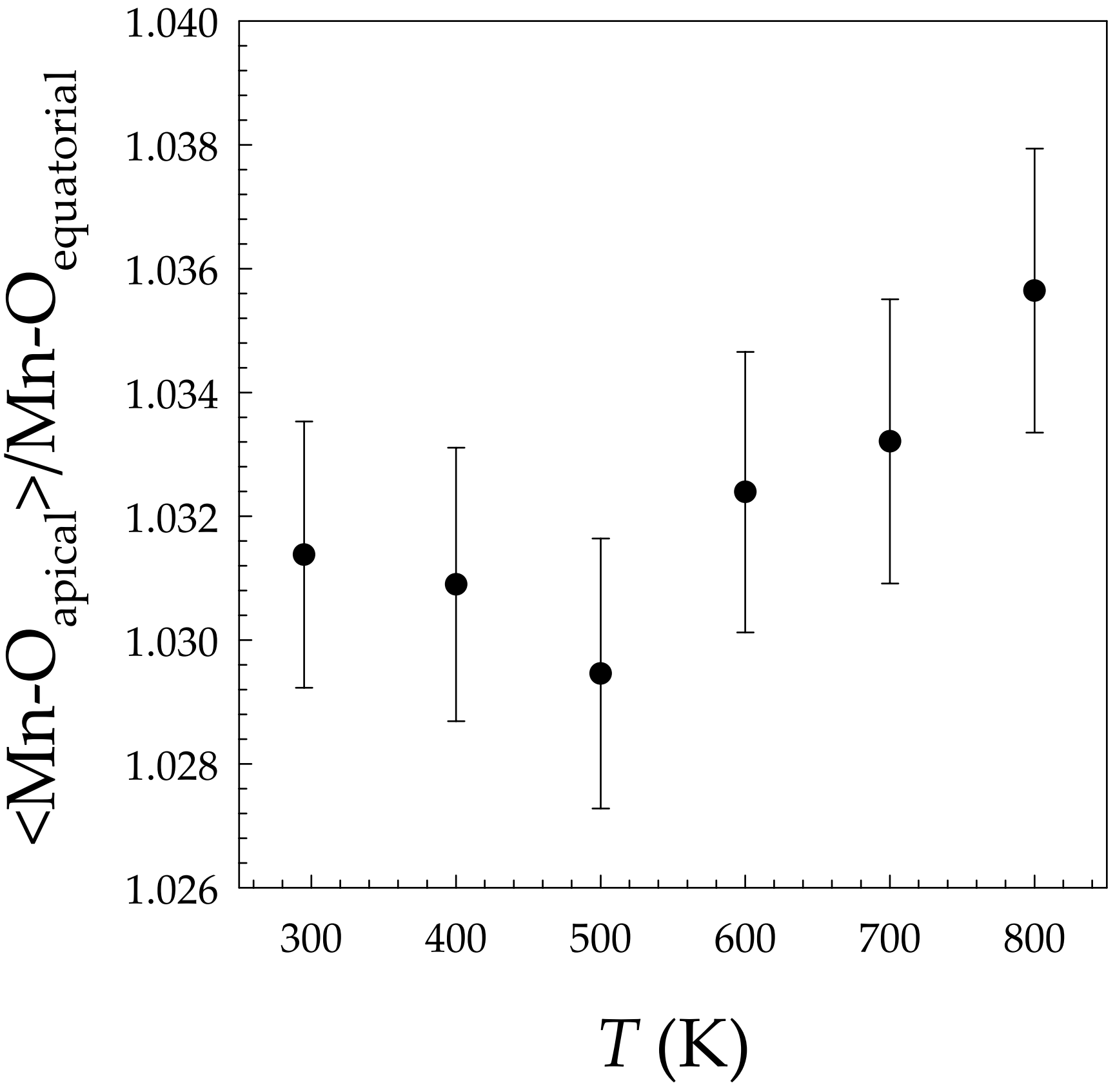

**Figure 5**

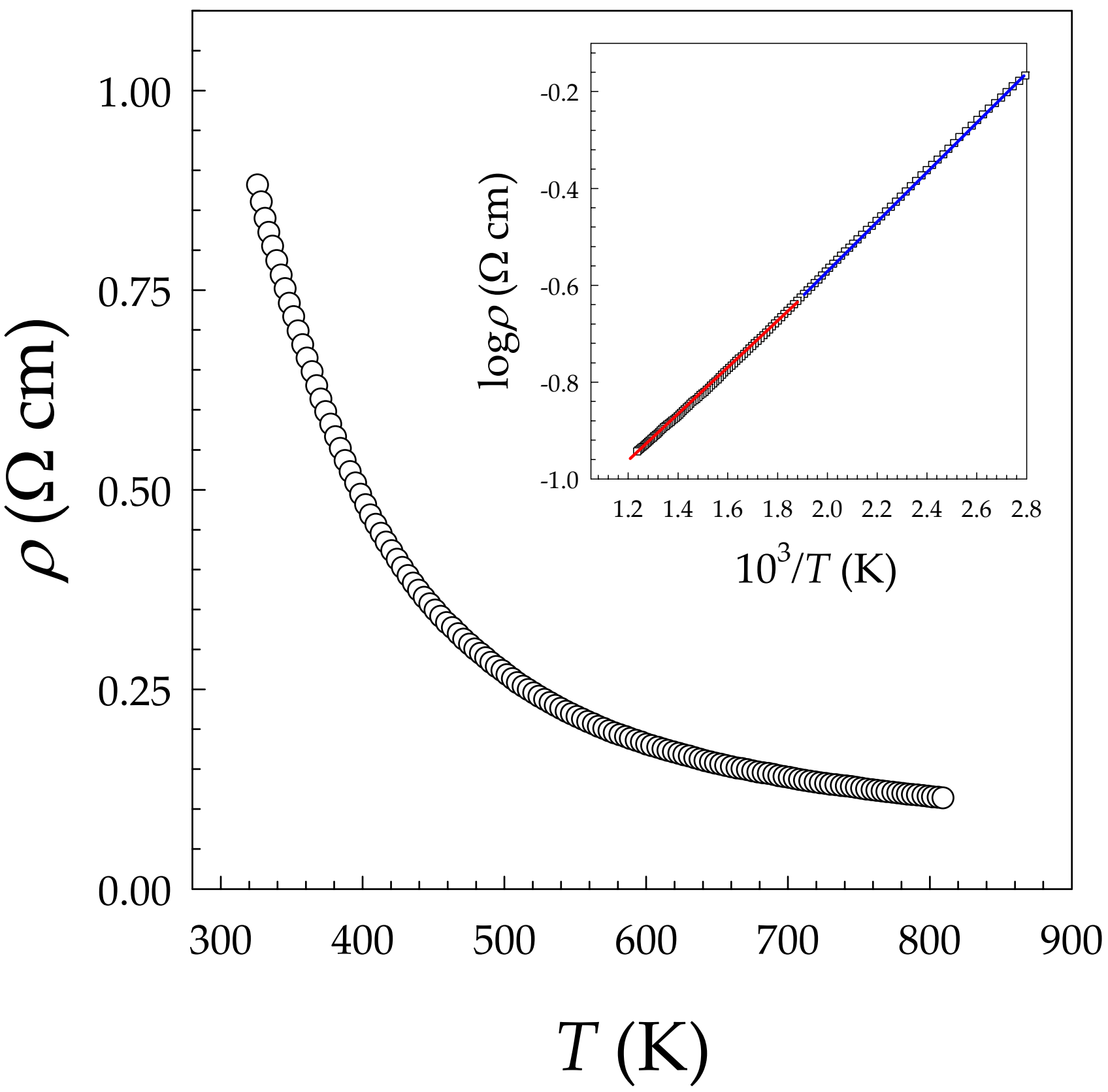

**Figure 6**